\documentclass[aps,prb,twocolumn,floatfix,floats,showpacs,superscriptaddress]{revtex4}
\usepackage{graphicx,latexsym}
\usepackage{dcolumn}
\usepackage{amsmath}

\usepackage{color}

\def\fig#1{Fig.~\ref{#1}}

\def\be{\begin{equation}}
\def\ee{\end{equation}}
\def\bea{\begin{eqnarray}}
\def\eea{\end{eqnarray}}

\begin{document}
\title {Transport spectroscopy in a time-modulated open quantum dot}

\author{C. S. Tang}
\email{cstang@phys.cts.nthu.edu.tw} \affiliation{Physics Division,
National Center for Theoretical Sciences, P.O. Box 2-131, Hsinchu
30013, Taiwan, ROC}
\author{Y. H. Tan}
\affiliation{Department of Electrophysics, National Chiao-Tung
University, Hsinchu 30010,Taiwan, ROC}
\author{C. S. Chu}
\affiliation{Department of Electrophysics, National Chiao-Tung
University, Hsinchu 30010,Taiwan, ROC}

\begin{abstract}
We have investigated the time-modulated coherent quantum transport
phenomena in a ballistic open quantum dot. The conductance $G$ and
the electron dwell time in the dots are calculated by a
time-dependent mode-matching method. Under high-frequency
modulation, the traversing electrons are found to exhibit three
types of resonant scatterings. They are intersideband scatterings:
into quasibound states in the dots, into true bound states in the
dots, and into quasibound states just beneath the subband
threshold in the leads.  Dip structures or fano structures in $G$
are their signatures. Our results show structures due to
2$\hbar\omega$ intersideband processes. At the above scattering
resonances, we have estimated, according to our dwell time
calculation, the number of round-trip scatterings that the
traversing electrons undertake between the two dot openings.
\end{abstract}



\maketitle

\section{Introduction}
\label{sec:int}

In the past decade, quantum transport phenomena in open quantum
dots has received much
attention.~\cite{Mar92,Chang94,Chan95,Persson95,Keller96,Wang96,
Akis97,Zoz98,Akis98,Bird99,Moura02}  The open quantum dot,
consisting of a submicron sized cavity connecting via point
contact leads to two end-electrodes, has become an important
device for the investigation of phase coherent processes and their
various mechanisms.   The size of the dot and the width of the
leads can be controlled by split-gates.  In high electron mobility
samples, and at sufficient low temperatures, the phase coherent
length may well exceed the dimension of the device, allowing
electrons to remain coherent while traversing the dot.

Meanwhile, there are growing interest in the high-frequency
responses of mesoscopic nanostructures. The time-modulated fields
invoked are either high-frequency electromagnetic
fields\cite{hek91,fen93,gor94,maa96,
chu96,wag97,tang99,Reichl00,tang00} or time-modulated
potentials.~\cite{bag92,tang96,tang98,Reichl99,tang01}  A number
of theoretical approaches have been developed to explore quantum
transport under such time-modulated fields.  WKB approximation was
employed in the study of photovoltaic effect~\cite{hek91} and
photon-assisted quantum transport.~\cite{fen93}  A mode-matching
method was developed for delta-profile~\cite{bag92} as well as
finite-range profile time-modulated potentials.~\cite{tang96}
Extension of this method to time-dependent field, represented by a
vector potential $\vec{A}(t)$, was carried out by either
neglecting~\cite{maa96} or including~\cite{tang99} the
contribution of $\vec{A}(t)^{2}$ term. This mode-matching method
was further extended to accommodate spatial inhomogeneity. The
time-modulated field is divided into piece-wise potentials
connected by either transfer matrices~\cite{wag97} or scattering
matrices.~\cite{tang98} Recently, this latter approach has been
applied to study a mechanism of nonadiabatic quantum
pumping.~\cite{tang01}  This pumping mechanism is due to
resonances resulted from coherent inelastic scattering that
requires simultaneous changes in both the energy and momentum of
the traversing electron, by respectively, $\hbar\Omega$ and
$\hbar{\it K\/}$. Here $\Omega$ and ${\it K\/}$ characterize,
respectively, the temporal and the spatial variation of the
modulation field. Encouraged by the success of the time-modulated
mode-matching method, we opt to apply the method to the very
interesting case of time-modulated quantum dots.

In the absence of a time-modulated field, transmission of
electrons through a quantum dot already shows resonance
structures. For the case of a weakly coupled dot --- dot in which
electrons are separated from the connecting leads by tunneling
barriers --- the resonance peaks in the transmission are due to
the alignment of the incident electron energy with the
quasi-bound-state (QBS) levels in the
dot.~\cite{Wang92,Tarucha96,Kouwenhoven97}  Interestingly, QBSs of
similar nature still exist in the case of an open quantum dot
--- where tunneling barriers between the dot and the lead are
absent. These QBSs, again, give rise to resonances in the
transmission. However, dip structures, rather than peaks, become
the signatures for the resonances.

When acted upon by a time-modulated potential, the transmission of
a weakly-coupled dot was found to exhibit additional resonance
peaks: peaks associated with ac sidebands.~\cite{Oosterkamp97}
This is due to the alignment, albeit shifted by $n\hbar\omega$, of
the incident electron energy with the QBS levels in the dot. Other
features found in a time-modulated weakly coupled dot are
photon-assisted tunneling,~\cite{Buttiker96} electron
pumps,~\cite{Geerlings} and phase breaking.~\cite{Anatram}

Recently, an open quantum dot, acted upon by a transversely
polarized electromagnetic field, and connected adiabatically to
the connecting leads, has been considered.~\cite{Gorelik97,
Gorelik01} The adiabatic dot-lead connections allow an electron
mode in the lead to evolve into an electron mode in the dot. Thus,
situations occur when an electron in the lower mode in the dot can
exit the dot without reflection, while an electron at the same
energy, but in a higher mode in the dot, is trapped inside it. As
such, inter-mode transitions between the above two modes in the
dot, as induced by the transversely polarized electromagnetic
field, were found to lead to giant mesoscopic conductance
fluctuations~\cite{Gorelik97} and microwave-induced resonant
blocking in a mesoscopic channel.~\cite{Gorelik01}

The adiabaticity of the dot-lead connection holds for large
quantum dots. But as the sizes of the open quantum dot shrink and
approach the realm of the Fermi wavelength, the dot-lead
connections could no longer remain adiabatic. More recent
experimental findings in dc transport in open quantum dots ---
that transport occurs through individual eigenstates of the
corresponding closed dot;~\cite{Zozou99} and that the conductance
oscillations correlate to the recurrence of specific groups of
wave function scars in the dot~\cite{Bird99} --- indicate
unequivocally that inter-mode scattering, and backscattering are
present, respectively, at the dot-lead connections. The effect of
impurity should play no role here because of the high mobility of
the sample used in these experiments.

Therefore, in this work, we consider a time-modulated open quantum
dot with non-adiabatic dot-lead connections. We calculate the dc
conductance $G$ of a time-modulated open quantum dot and the dwell
time $\tau_{\rm d}$ of the traversing electron in the dot. We have
analyzed the resonance structures in $G$ associated with the
time-modulation and are able to categorize them according to their
respective dynamical processes involved. Of these three resonance
types, one is analogous to that found in time-modulated weakly
coupled dots. It is associated with the alignment of the incident
electron energy with that of the ac sidebands of the QBSs inside
the open dot. The second type is associated with the coherent
inelastic scattering of the traversing electron into the true
bound state in the open dot---bound state which energy is lower
than the threshold energy of the leads. The third type of
resonance structures is most unexpected. It is associated with the
coherent inelastic scattering of the traversing electron into the
QBS in the lead---with energy just beneath the threshold energy of
the lead. Also, from the dwell time $\tau_{\rm d}$, we estimate
the number of scatterings that occur in the dot as the resonance
structures establish themselves. In all, our results demonstrate
the potential of establishing quantum transport as a spectroscopic
probe for the QBSs and true bound states in the open dot---and
possibly in other mesoscopic structures---through the coupling of
a time-modulated field to the system.

In section~\ref{sec:model}, we present our theoretical method for
the calculation of $G$ and $\tau_{\rm d}$. The numerical results
are presented and discussed in Sec.~\ref{sec:numerical}. Finally,
in Sec.~\ref{sec:conclusion}, we present our conclusions.

\section{Model and method}
\label{sec:model}

\begin{figure}[bp]
\includegraphics[width = 0.48\textwidth]{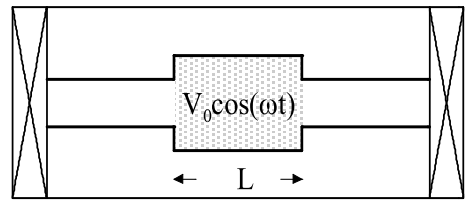}
\caption{Schematic illustration of an open quantum dot which is
acted upon by a gate-induced time-modulation with two leads
connecting adiabatically  to two end-electrodes.} \label{fig1}
\end{figure}

The system under investigation is sketched (top view) in \fig{fig1},
where the shaded area denotes the region acted upon by a
time-modulated potential.  The dot we consider has physical
parameters typical to that in high mobility two-dimensional electron
gas (2DEG), formed in an AlGaAs-GaAs heterostructure. As such,
mobility $\mu_e\sim 10^6 {\rm cm^2/Vs}$, mean free path $l\sim 1
\mu{\rm m}$ at sufficient low temperatures, and dots with submicron
dot sizes would be in the ballistic regime. The Hamiltonian is given
by
\begin{equation}
H=-\frac{\hbar^2}{2m}\left[\frac{\partial^2}{\partial
x^{2}}+\frac{\partial^2}{\partial^2 y}\right]+  V_c(x,y) + V(x,t),
\label{Eqn:H}
\end{equation}
where $V_{c}(x,y)$ is the confinement potential, chosen to be of
hard-wall type, that defines the dot and the leads. It is given by
$V_c = 0$ if $|y|<W_1/2$ and $|x|> L/2$; $V_c = 0$ if $|y|<W_2/2$
and $|x|< L/2$; and $V_c = \infty$ if otherwise. Here
$W_{1},W_{2}$ are, respectively, widths of the lead and the dot.
The time-modulated potential
$$V(x,t)=V_0\cos\left(\omega t\right)\, \Theta \left(
L/2-\left|x\right|\right)$$ acts only upon the dot.

For the sake of convenience, the physical quantities that appear
in the following equations are dimensionless: with energy unit
$E^*=E_F=\hbar^2k^2_F/2m$, wave vector unit $k^* = k_F$, length
unit $a^*=1/k_F$, time unit $t^* = \hbar/E_F$, and frequency unit
$\omega^*=1/t^*$. The scattering wavefunction for an electron
incident upon the dot from the $l$-th channel in the left-lead, is
of the form

\begin{widetext}
\begin{equation*}
\psi_l(x,y,t)= \chi_l(y)e^{ik_l(0)x}e^{-i\mu t} +\sum\limits_{n'}
\sum\limits_{m'} \chi_{n'}(y) r_{n'l}(m') \exp \left[-ik_{n'}(m')x
-i(\mu+m'\omega)t \right],\, {\rm if} \, x< -L/2,
\end{equation*}
\begin{equation}
\psi_l(x,y,t)= \sum\limits_{k'} \phi_{k'}(y) \int d\epsilon
[\widetilde{A}_{k'l} (\epsilon)e^{i\beta_{k'}(\epsilon)x}\\
\, +\widetilde{B}_{k' l}(\epsilon)e^{-i\beta_{k'}(\epsilon)x}]
\exp \left[ -i\epsilon t -i \frac{V_0}{\omega}\sin\omega t
\right],\, {\rm if}\, |x|< L/2,
\end{equation}
\begin{equation*}
\psi_l(x,y,t)= \sum\limits_{n'} \sum\limits_{m'} \chi_{n'}(y)
t_{n'l}(m') \exp\left[ ik_{n'}(m')x -i(\mu+m'\omega)t \right],\,
{\rm if} \, x>L/2,
\end{equation*}
\end{widetext}
where the subscripts $n'$ and $k'$ are the subband indices in,
respectively, the leads and the dot, and $m'$ is the sideband
index. In addition,
$k_l(m')=\left(\mu+m'\omega-(l\pi/W_1)^2\right)^{1/2}$ and
$\beta_{k'}(m')=\left(\mu+m'\omega-(k'\pi/W_2)^2\right)^{1/2}$
denote, respectively, the wave vectors in the lead and the dot.
The normalized transverse subband states are $\chi_{l}(y) =
(2/W_1)^{1/2}\sin\left[l\pi(y/W_1 + 1/2)\right]$ and $\phi_{k'}(y)
= (2/W_2)^{1/2}\sin\left[k'\pi(y/W_2+1/2)\right]$.

The matching of the wave functions at the openings of the dot, and
at all times, requires the coefficients in the dot to have the
form \[\widetilde{\cal F}_{k' l}(\epsilon)=\sum\limits_{m'}{\cal
F}_{k'l}(m')\delta(\epsilon-\mu-m'\omega),\] where
$\widetilde{\cal F}_{k'l}(\epsilon)$ refers to either $\widetilde
A_{k'l}(\epsilon)$ or $\widetilde B_{k'l}(\epsilon)$. Performing
the matching, and after some algebra, we obtain
\begin{widetext}
\begin{eqnarray}
\lefteqn{
 \left[ A_{kl}(m)
\exp\left(-i\beta_{k}(m){L\over 2}\right)
+B_{kl}(m)\exp\left(i\beta_{k}(m){L\over2}\right) \right]}\nonumber \\
&=& \sum\limits_{n'}\sum\limits_{m'}
J_{m'-m}\left(\frac{V_0}{\omega}\right) \left[ a_{lk}
\exp\left(-ik_l(m'){L\over 2}\right)\delta_{n' l}\delta_{m'0} +
a_{n'k}r_{n'l}(m') \exp\left(ik_{n'}(m'){L\over 2}\right) \right],
\label{Eqn:1}
\end{eqnarray}
\begin{eqnarray}
\lefteqn{ \left[A_{kl}(m) \exp\left(i\beta_{k}(m){L\over
2}\right)+ B_{kl}(m)\exp\left(-i\beta_{k}(m){L\over
2}\right)\right] }
\nonumber \\
&=& \sum\limits_{n'}\sum\limits_{m'}
J_{m'-m}\left(\frac{V_0}{\omega}\right) \left[a_{n'k}\,
t_{n'l}(m') \exp\left(ik_{n'}(m'){L\over2}\right) \right] ,
\label{Eqn:2}
\end{eqnarray}
\begin{eqnarray}
\lefteqn{ \sum\limits_{k'}a_{nk'}\beta_{k'}(m)
\left[A_{k'l}(m)
\exp\left(-i\beta_{k'}(m){L\over 2}\right)-
B_{k'l}(m)\exp\left(i\beta_{k'}(m){L\over 2}\right)\right]
}\nonumber \\
&=& \sum\limits_{m'} J_{m'-m}\left(\frac{V_0}{\omega}\right)
k_n(m') \left[ \delta_{nl}\delta_{m'0}
\exp\left(-ik_l(m'){L\over2}\right) - r_{nl}(m')
\exp\left(ik_{n'}(m'){L\over2}\right) \right] , \label{Eqn:3}
\end{eqnarray}
and
\begin{eqnarray}
\lefteqn{ \sum\limits_{k'}a_{nk'}\beta_{k'}(m)\left[A_{k'l}(m)
\exp\left(i\beta_{k'}(m){L\over 2}\right)-
B_{k'l}(m)\exp\left(-i\beta_{k'}(m){L\over 2}\right)\right]
}\nonumber \\
&=& \sum\limits_{m'}
J_{m'-m}\left(\frac{V_0}{\omega}\right)\,k_{n}(m') t_{nl}(m')
\exp\left(ik_{n}(m'){L\over 2}\right) , \label{Eqn:4}
\end{eqnarray}
\end{widetext}
where Eqs.(5) and (6) are obtained from matching the derivatives
of the wave functions. The overlapping integral $a_{lk}$ of the
transverse subband states is given by
\begin{equation}
a_{lk} = \int_{-W_1 /2}^{W_1 /2}\chi_l(y) \phi_k(y)dy ,
\label{Eqn:a}
\end{equation}
and the identity $\exp\left( iz \sin\omega t\right) = \sum_p
J_p(z)\exp \left(ip\,\omega t\right)$ has been invoked. We have
solved Eqs.(\ref{Eqn:1})-(\ref{Eqn:4}) for the coefficients
$A_{kl}(m)$, $B_{kl}(m)$, $r_{n'l}(m')$ and $t_{n'l}(m')$.
Furthermore, we note that the sole appearance of $V_{0}$ in
Eqs.(\ref{Eqn:1})-(\ref{Eqn:4}) is in the form $V_{0}/\omega$, and
as an argument of the Bessel functions $J_{m}$. This shows a
general trend that the effect of the time-modulated potential
decreases with the raising of the frequency $\omega$.

In the low drain-source bias regime, the dc conductance is given
by
\begin{equation}
G=\frac{2e^2}{h}\sum_{l=1}^{N} T_l
\end{equation}
where $N$ denotes the number of propagating channels in the leads.
The current transmission coefficient $T_l$ for an electron
incident from the $l$-th channel in the lead is
\begin{equation}
T_l=\sum\limits_{n'}\sum\limits_{m'}\frac{k_{n'}(m')}{k_l(0)}
\left|t_{n'l}(m')\right|^2.
\end{equation}
The current reflection coefficient $R_l$ has a similar form, and
the current conservation condition $T_l+R_l=1$ is used to check on
our numerical accuracy.

The stationary dwell time within one-dimensional system was well
defined.~\cite{Smith} However, in a multi-channel system such as
open quantum dots, we should consider not only the probability of
finding the particle in the dot but also that due to evanescent
states in the vicinity of the dot. Hence, we define the dwell time
as
\begin{equation}
\label{dt} \tau_d = \frac{\int\int_{\cal A^\prime}
\langle|\psi(x,y,t)|^2\rangle_{\rm t.a.}dxdy}{\int dy\,j_{\rm
inc}}
\end{equation}
where $j_{\rm inc}$ denotes the incident electron flux. The
subscript ${\rm t.a.}$ denotes time average.  Here we note that
the integral in the numerator and its region of interest ${\cal
A^\prime}$ include not only the region inside the quantum dot
(region II), but also the evanescent modes on both the left-hand
side (region I) and the right-hand side (region III) of the dot.
Hence, the  time-averaged probability density in the numerator of
Eq. (\ref{dt}) can be separated into three integrals, expressed
explicitly as
\begin{eqnarray}\label{I}
\lefteqn{\int\int_{\rm I} dxdy \langle|\psi(x,y,t)|^2\rangle_{\rm t.a.}}\nonumber \\
&=&\int_{-\infty}^{-L/2} dx
\sum\limits_{n^\prime}\sum\limits_{m'}|r_{n'l}(m')|^2e^{2\kappa_{n'}(m')
x}\, ,
\end{eqnarray}
\begin{eqnarray}\label{II}
\lefteqn{\int\int_{\rm II} dxdy \langle|\psi(x,y,t)|^2\rangle_{\rm t.a.}}\nonumber \\
&=&\int_{-L/2}^{L/2}dx\sum\limits_{k'}\sum\limits_{m'}
\left\{  A_{k'l}(m')A^*_{k'l}(m') \right. \nonumber \\
&&\times \exp\left[i\left( \beta_{k'}(m') -\beta^*_{k'}(m') \right) x \right]\nonumber \\
&&+ B_{k'l}(m')B^*_{k'l}(m')e^{-i\left(\beta_{k'}(m') -\beta^*_{k'}(m')\right)x} \nonumber \\
&&+ \left. 2{\rm Re}
\left[A_{k'l}(m')B^*_{k'l}(m')e^{i\left(\beta_{k'}(m') +
\beta^*_{k'}(m') \right)x}\right]\right\} , \nonumber \\ & &
\end{eqnarray}
and
\begin{eqnarray}\label{III}
\lefteqn{\int\int_{\rm III} dxdy \langle|\psi(x,y,t)|^2\rangle_{\rm t.a.}}\nonumber \\
&=&\int_{L/2}^{\infty} dx
\sum\limits_{n'}\sum\limits_{m'}|t_{n'l}(m')|^2\exp\left(-2\kappa_{n'}(m')
x\right)
\end{eqnarray}
where the indices $n'$ and $m'$ in the summation for regions
${\rm I}$ and ${\rm III}$ include only the evanescent waves, and
$\kappa_{n^\prime}(m)=-ik_{n'}(m)$.  Substituting
Eqs.~(\ref{I})-(\ref{III}) into Eq.~(\ref{dt}), we obtain the
average dwell time of electrons in the open quantum dot system.

\section{RESULTS AND DISCUSSION}
\label{sec:numerical}

In this section, we present our numerical examples for exploring
the time-modulated effects on the quantum transport in open
quantum dots --- the conductance and the dwell time versus the
incident electron energy $\mu$.   In the following, we choose
energy unit $E^*=9~{\rm meV}$, length unit $a^*=8$ nm, unit of
angular frequency $\omega^*=13.6$ Trad/sec, and the effective mass
$m^*=0.067m_e$ where $m_e$ is the free electron mass of an
electron. The geometric parameters are chosen such that the width
$W_1=10$ $(\cong 80 {\rm nm})$ for the leads, and the width
$W_2=20$ $(\cong 160 {\rm nm})$ and the length $L=30$ $(\cong 240
{\rm nm})$ for the open quantum dot, which is typical for current
experimental fabrication.  It is convenient to define $X^2 \equiv
\mu/\varepsilon_1$ where $\varepsilon_1 = \left(\pi/W_1\right)^2$
is the first transverse subband level in the leads, then the
integral values of $X$ stand for the number of occupied subbands
in the leads.

In the absence of time-modulation, the quantum states of the open
quantum dot associate closely with the bound states of the
corresponding closed dot with the same geometry.  For a closed dot
with length $L$ and width $W_2$, the bound state energy $E_{\rm
b.s.}= (n_x \pi/L)^2 + (n_y\pi/W_2)^2$ is labelled by a pair of
quantum number $(n_x,n_y)$. Then we may obtain the rescaled bound
state levels $\underline{E}(n_x,n_y)\equiv E_{\rm
b.s.}/\varepsilon_1 = n_x^2/9+n_y^2/4$ inside the closed dot. For
an open dot, the energy spectrum consists of both true bound
states and QBSs corresponding either to electrons with energy
$\mu$ less than or higher than the threshold energy
$\varepsilon_1$ in the lead. As such, there are only two possible
true bound states in the open dot: $E_{B1}\simeq
\underline{E}(1,1)= 0.361$ and $E_{B2}\simeq \underline{E}(2,1)=
0.694$.

\begin{figure}[bp]
\includegraphics[width = 0.48\textwidth]{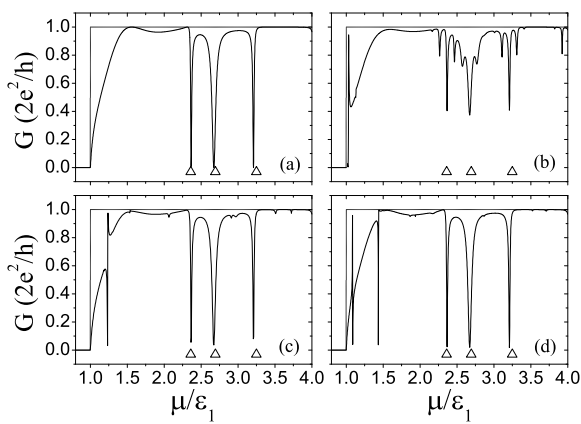}
\caption{Energy dependence of the quantum dot conductance $G$ in the
lowest subband as a function of incident electron energy $\mu$, in
units of $\varepsilon_1$, for cases of: (a) no external modulation;
(b)-(d)  modulation amplitude $V_0 = 0.1 \varepsilon_1$, with
angular frequencies (b)$\omega = 0.1\varepsilon_1$,
(c)$0.3\varepsilon_1$, and (d)$0.5 \varepsilon_1$. }
 \label{fig2}
\end{figure}

\begin{figure}[bp]
\includegraphics[width = 0.48\textwidth]{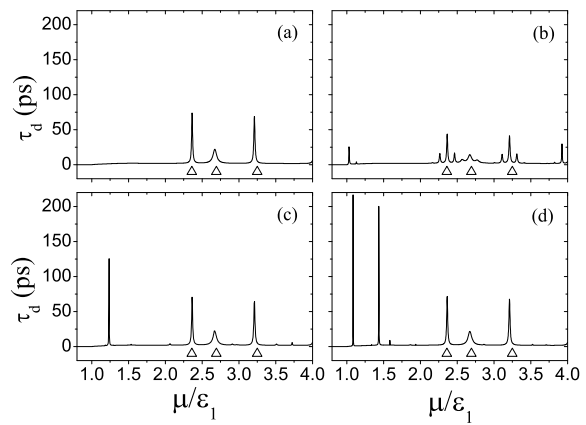}
\caption{The dwell time $\tau_d$ of the traversing electron is
plotted as a function of incident electron energy $\mu$ in units of
$\varepsilon_1$.  The parameters are the same as in
\fig{fig2}(a)-(d). } \label{fig3}
\end{figure}

In \fig{fig2}, the conductance characteristic is studied as a
function of incident electron energy: (a)  in the absence of
time-modulation, as a comparative reference; and (b)-(d) in the
presence of time-modulation with angular frequencies
$\omega/\varepsilon_1 = 0.1$, $0.3$, and $0.5$, respectively, which
also correspond to frequencies $f = \omega/2\pi\simeq 21.4$, $64.2$,
and $107$ GHz. This frequency range is typical for current
experiments.~\cite{Qin01} In addition, the modulation amplitude is
chosen to be $V_0 = 0.1 \varepsilon_1$ $\left(\simeq 0.09~{\rm
meV}\right)$. There are three dip structures common to all four
plots.  These dip structures occur at energies $\mu/\varepsilon_1 =
2.364\, (E_{Q1})$, $2.672\, (E_{Q2})$, and $3.208\, (E_{Q3})$. These
are associated with the alignment of the incident electron energy
with that of the QBS levels inside the open dot, as is indicated by
the open triangle symbols locating the closed dot
$\underline{E}(1,3)= 2.361$, $\underline{E}(2,3)= 2.694$, and
$\underline{E}(3,3)= 3.250$.   The corresponding dwell time of these
QBSs are, respectively,  $\tau_d \simeq 73.5$, $26.7$, and $69.0$
ps, as shown in \fig{fig3}(a).  These dwell time peak structures
confirm the resonant nature of the states inside the dot.

Another interesting feature in \fig{fig2}(b) is the side-dip
structures around the QBSs $E_{Qi}$, which is associated with
electrons at incident energy $\mu$ that are able to make $m$-photon
intersideband transitions into the $i$-th QBS level. The condition
is
\begin{equation}
\mu +m\hbar\omega =E_{Qi},
\end{equation}
where the positive $m$ and negative $m$ indicate, respectively,
the  absorption and emission of $m$-photons.  Hence $m=-1$
side-dips are at $\mu/\varepsilon_1 = 2.465$, $2.770$, $3.309$;
$m=+1$ side-dips are at $\mu/\varepsilon_1 = 2.165$, $2.574$,
$3.110$; and $m=+2$ side-dips are at  $\mu/\varepsilon_1 = 2.266$,
$3.008$.  The $m=-2$ process in the vicinity of the $E_{Q3}$ state
is barely identified and is at
 $\mu/\varepsilon_1 = 3.411$.

Two additional types of intersideband transition mechanisms are
found in the low energy regime in \fig{fig2}.  As is shown in
\fig{fig2}(d), whereas the frequency $\omega = 0.5\varepsilon_1$ is
high enough, the electrons with $\mu/\varepsilon_1 = 1.087$ may emit
$\hbar\omega$ and make
 transitions into $E_{B2}$ --- the true bound state in
the dot. The electron dwell time of this structure is $\tau_d \cong
216.6$ ps, see \fig{fig3}(d). In addition, electrons may also emit
photons to make transitions into a QBS formed at energy just beneath
a subband threshold in the lead.  This mechanism is identified to be
the fano structures at $\mu/\varepsilon_1 = 1.028$, $1.233$, and
$1.435$, as shown in \fig{fig2}(b)-(d), respectively, where
$(\mu-\omega)/\varepsilon_1$ is close to $1$ from below.
Correspondingly, the dwell time of these structures are $\tau_d
\simeq 23.0$, $125.5$, and $200.2$ ps, see \fig{fig3}(b)-(d). More
precisely, these structures correspond to electrons that emit
$\hbar\omega$ to $\mu/\varepsilon_1 = 0.93$ and being trapped
temporarily to form QBSs in the lead.

\begin{figure}[bp]
\includegraphics[width = 0.48\textwidth]{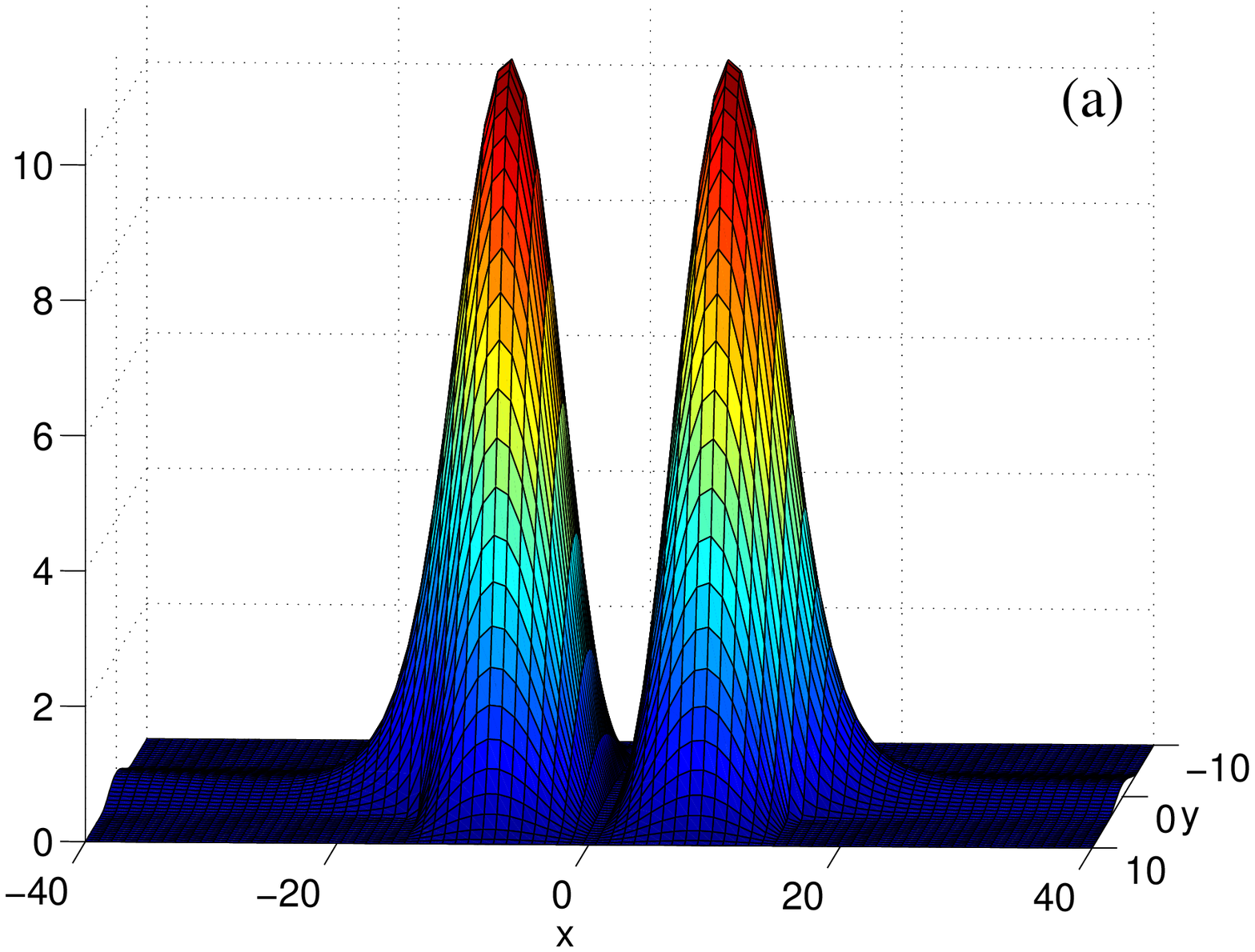}
\includegraphics[width = 0.48\textwidth]{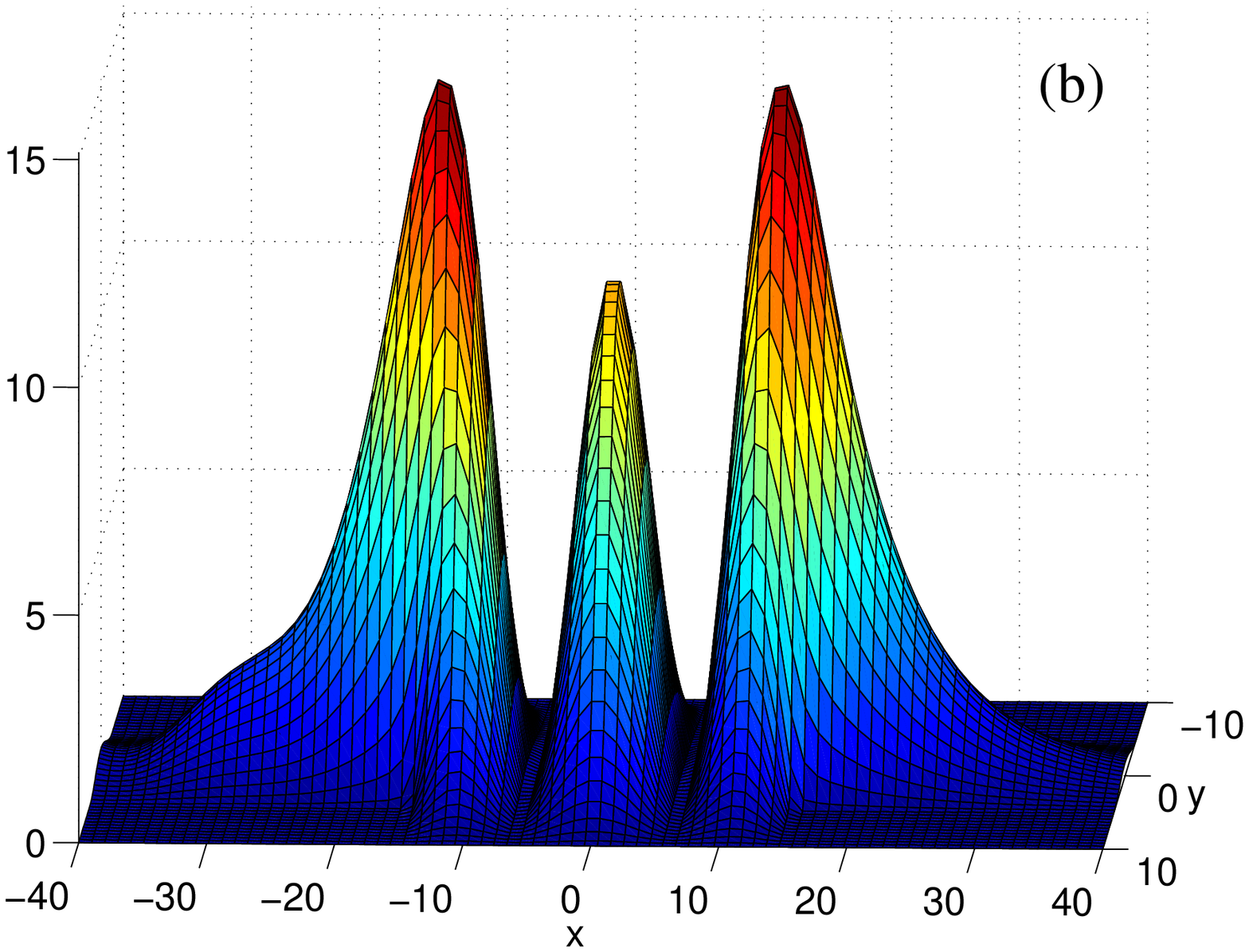}
\caption{(color). The spatial dependence of the time-averaged
electron probability density: (a)$\mu = 1.0854 \varepsilon_1$ and
(b)$\mu=1.4353 \varepsilon_1$. Other parameters are the same as in
\fig{fig2}(d). The edges of the dot are at $L=\pm 15$} \label{fig4}
\end{figure}
To provide further evidence for the above two transition mechanisms,
we plot, in \fig{fig4}, the spatial dependence of the time-averaged
probability density.  The parameters are chosen to be the same as in
\fig{fig2}(d).  When the electron incident at energy $\mu = 1.0854
\varepsilon_1$, as shown in \fig{fig4}(a), we see that the
time-averaged electron probability concentrates entirely within the
dot and is like a (2,1) state. This support the fact that the
electron being trapped in a true bound state in the dot. Second,
when the electron incident at energy $\mu=1.4353\varepsilon_1$, the
electron probability has long exponential tails extending into the
leads. The edges of the dot are at $x=\pm 15$.  This demonstrates
that the electron has made an intersideband transition, by emitting
one photon, into a QBS in the lead which energy is just below the
threshold energy.  We have also checked that similar process can be
found even in a time-modulated one dimensional quantum well
connecting to leads. A traversing electron can make intersideband
transitions into QBSs in the leads or into true bound states in the
well.~\cite{tang-unp}

To better appreciate the meaning of the dwell time, we define the
number of round-trip scatterings $N_{\rm sc}$ undertaken by the
traversing electron.  It is the ratio of the dwell time $\tau_d$ to
the ballistic time $\tau_b$ it takes the electron to go between the
two dot-openings. The ballistic time for electrons traversing
through the quantum dot is simply $\tau_b \sim L/v_e$ where $v_e$
denotes the electron velocity, given by $v_e = \hbar k_x/m$.  We
consider the electron incident in the lowest subband and then the
electron ballistic time is given by $\tau_b = L/ \left(\mu -
\varepsilon_1 \right)^{1/2}$ in units of $t^*$. Hence, in
\fig{fig3}(d), the main peaks at $\mu/\varepsilon_1 = 1.085$,
$1.435$, $2.365$, $2.673$, and $3.210$ correspond to the ballistic
times $\tau_b \simeq 24.0$, $16.1$, $6.01$, $5.43$, and $4.72$ ps,
respectively. The corresponding $\tau_{d}$'s are, respectively,
$216.6$, $200.2$, $71.8$, $21.8$, and $67.9$. Therefore, we obtain
the number of round-trip scatterings $N_{\rm sc}\equiv \tau_d /
2\tau_b \sim 4.5$, $6.2$, $6.0$, $2.0$, and $7.2$, respectively. In
light of the above analysis, we can see that two round-trip times
are already sufficient to form a significant QBS level inside the
open dot.  The estimation for $\tau_b$ could be improved by
considering the effective electron velocity in the dot, rather than
in the lead. But we expect $N_{\rm sc}$ to remain of the same order
of magnitude as what we have shown here. Moreover, the $N_{\rm sc}$
obtained here is the lower bound to any such improved estimation.

\begin{figure}[bp]
\includegraphics[width = 0.48\textwidth]{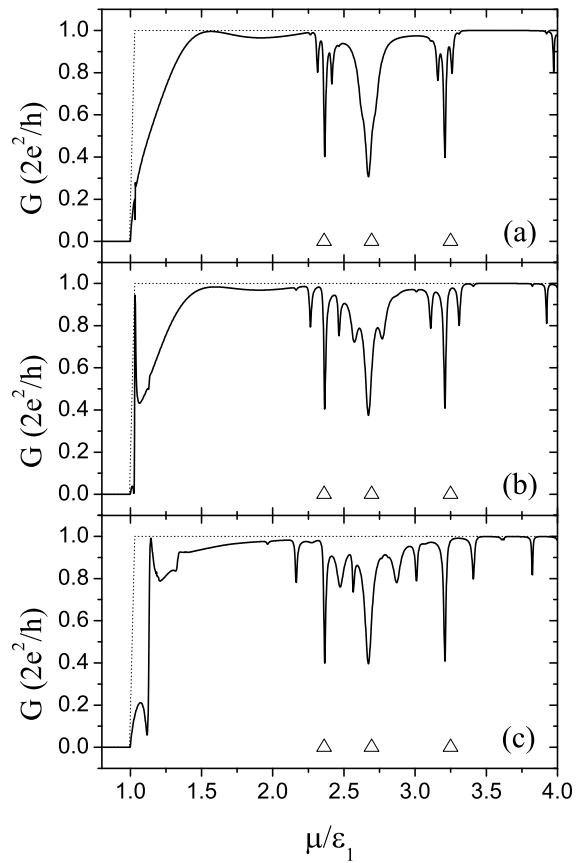}
\caption{Conductance is plotted as a function of electron energy
$\mu$ for a fixed $V_0/\omega = 1.0$. The values of $V_0$ (or
$\omega$) are (a)$0.05\varepsilon_1$, (b)$0.1\varepsilon_1$, and
(c)$0.2\varepsilon_1$, respectively.} \label{fig5}
\end{figure}
It is known~\cite{tang96} that the strength of the time-modulated
potential depends on the ratio of  $V_0$ to $\omega$. As a result,
for a given amplitude $V_0$, the sideband dip features are
suppressed with the increasing of the modulation frequency $\omega$.
This is illustrated in \fig{fig2}(c) and (d).  On the other hand, if
we fix the $V_0/\omega = 1$ in \fig{fig5}, and choose the modulation
amplitude to be $V_0/\varepsilon_1 = 0.05$, $0.1$, and $0.2$ for
\fig{fig5}(a)-(c), respectively, we can see the side-dips due to
$2\hbar\omega$ intersideband processes in all the figures. This
assures us that the $V_0/\omega$ is an important index for photon
absorption and emission processes. Furthermore, the QBS levels that
associate with fewer number of round-trip scatterings may be merged
with its nearby sidebands, forming a broadened dip structure: such
as the wide-dip structure at $\mu/\varepsilon_1 = 2.672$ in
\fig{fig5}(a). In the low energy regime, again, electrons are able
to undertake one-photon (or two-photon) emission processes into a
subband threshold in the lead. This one- (or two-) photon mechanism
is demonstrated by small dip structures in $G$ at $\mu/\varepsilon_1
= 1.024$ (or $1.129$) in \fig{fig5}(b) and $\mu/\varepsilon_1 =
1.118$ (or $1.317$) in \fig{fig5}(c).

To conclude this section, we note in passing that despite of wide
dot openings, electrons traversing through the dot are still
effectively mediated by just a few  bound states of the
corresponding closed dot structure.  Our results show that the
conductance spectra for a time-modulated open dot show more
intersideband structures other than those associated with the
bound states of the corresponding close dot. We believe that these
mechanisms  should find their way of manifestation in the
time-modulated phenomena of other nanostructures.

\section{CONCLUSIONS}
\label{sec:conclusion}

In this work we have extended the time-dependent mode matching
approach to the study of quantum transport in open quantum dot
systems.  We have calculated the conductance, the  dwell time,
and the spatial distribution of the electron probability and
their dependence on the modulation amplitudes and frequencies.

In conclusion, we have shown three types of coherent inelastic
scatterings in a time-modulated open quantum dot. We have
demonstrated the potential of establishing quantum transport as a
spectroscopic probe for the QBSs and true bound states in an open
dot through the coupling of a time-modulated field to the system.

\begin{acknowledgments}
The authors wish to acknowledge the National Science Council of
the Republic of China for financially supporting this research
under Grant No. NSC90(91)-2119-M-007-004 (NCTS),
NSC90-2112-M-262-001 (CST), and NSC90-2112-M-009-044 (CSC).
Computational facilities supported by the National Center for
High-Performance Computing are gratefully acknowledged.
\end{acknowledgments}


\begin{references}

\bibitem{Mar92}  C.M. Marcus, A.J. Rimberg, R.M. Wetervelt,
                 P.F. Hopkins, and A.C. Gossard, Phys. Rev. Lett. {\bf
                 69}, 506 (1992)
\bibitem{Chang94}A.M. Chang, H.U. Baranger, L.N. Pfeiffer, and K.W. West,
                 Phys. Rev. Lett. {\bf 73}, 2111 (1994).
\bibitem{Chan95}H.I. Chan, R.M. Clarke, C.M. Marcus, K. Campman, and A.C.
                Gossard, Phys. Rev. Lett. {\bf 74}, 3876 (1995).
\bibitem{Persson95} M. Persson, J. Pettersson, B. von Sydow, P.E. Lindelof, A. Kristensen,
                  and K.-F. Berggren, Phys. Rev. B {\bf 52}, 8921 (1995).
\bibitem{Keller96}M.W. Keller, A. Mittal, J.W. Sleight, R.G. Wheeler, D.E.  Prober,
                  R.N. Sacks, and H. Shrtikmann, Phys. Rev. B {\bf 53}, R1 693 (1996).
\bibitem{Wang96} Y. Wang, N. Zhu, and J. Wang, Phys. Rev. B {\bf
                 53}, 16408 (1996).
\bibitem{Akis97} R. Akis, D.K. Ferry, and J.P. Bird, Phys. Rev. Lett.
                 {\bf 79} 123 (1997)
\bibitem{Zoz98}  I.V. Zozoulenko and T. Lundberg, Phys. Rev. Lett.
                 {\bf 81}, 1744 (1998).
\bibitem{Akis98} R. Akis, D.K. Ferry, and J.P. Bird, Phys. Rev. Lett.
                 {\bf 81}, 1745 (1998).
\bibitem{Bird99} J.P. Bird, R. Akis, D.K. Ferry, D. Vasileska, J. Cooper,
                 Y. Aoyagi, and T. Sugano, Phys. Rev. Lett. {\bf 82}, 4691
                 (1999).
\bibitem{Moura02}A.P.S. de Moura, Y.-C. Lai, R. Akis, J.P. Bird, and
                 D.K. Ferry, Phys. Rev. Lett. {\bf 88}, 6804 (2002).

\bibitem{hek91} F. Hekking, and Y.V. Nazarov, Phys. Rev. B {\bf 44},
                11506 (1991).
\bibitem{fen93} S. Feng and Q. Hu, Phys. Rev. B {\bf 48}, 5354 (1993).
\bibitem{gor94} L.Y. Gorelik, A. Grincwajg, V.Z. Kleiner, R.I.
                Shekhter, and M. Jonson, Phys. Rev. Lett. {\bf 73},
                2260 (1994).
\bibitem{maa96} F.A. Maa\o\ and L.Y. Gorelik, Phys. Rev. B
                {\bf 53}, 15885 (1996).
\bibitem{chu96} C.S. Chu and C.S. Tang, Solid State Commun.
                {\bf 97}, 119 (1996).
\bibitem{wag97} M. Wagner and W. Zwerger, Phys. Rev. B {\bf 55}, R10217 (1997).
\bibitem{tang99} C.S. Tang and C.S. Chu, Phys. Rev. B {\bf 60},
                 1830 (1999).
\bibitem{Reichl00} Wenjun Li and L.E. Reichl, Phys. Rev. B {\bf 62}, 8269 (2000).
\bibitem{tang00} C.S. Tang and C.S. Chu, Physica B {\bf 292}, 127 (2000).

\bibitem{bag92}
P.F. Bagwell and R.K. Lake, Phys. Rev. B {\bf 46}, 15329 (1992).
\bibitem{tang96}
C.S. Tang and C.S. Chu, Phys. Rev. B {\bf 53}, 4838 (1996).
\bibitem{tang98}
C.S. Tang and C.S. Chu, Physica B {\bf 254}, 178 (1998).
\bibitem{Reichl99}
Wenjun Li and L.E. Reichl, Phys. Rev. B {\bf 60}, 15732 (1999).
\bibitem{tang01}
C.S. Tang and C.S. Chu, Solid State Commun. {\bf 120}, 353 (2001).



\bibitem{Wang92}
J. Wang and H. Guo, Appl. Phys. Lett. {\bf 60}, 654 (1992).
\bibitem{Tarucha96}
S. Tarucha, D.G. Austing, T. Honda, R.J. van der Hage, and L.P.
Kouwenhoven, Phys. Rev. Lett. {\bf 77}, 3613 (1996).
\bibitem{Kouwenhoven97}
L.P. Kouwenhoven {\it et al.}, in {\it Mesoscopic Electron
Transport, Proceeding of a NATO Advanced Study Institute}, edited
by L.L. Sohn, L.P. Kouwenhoven, and G. Sch\"{o}n (Kluwer,
Dordrecht, 1997). Ser. E, Vol. 345.
\bibitem{Oosterkamp97}
T.H. Oosterkamp, L.P. Kouwenhoven, A.E.A. Kooten, N.C. van der
Vaart, and C.J.P.M. Harmans, Phys. Rev. Lett. {\bf 78}, 1536
(1997).
\bibitem{Buttiker96}
T. Christen and M. B\"{u}ttiker, Phys. Rev. B {\bf 53}, 2064
(1996).

\bibitem{Geerlings}
L.J. Geerlings, V.F. Anderegg, P.A.M. Holweg, J.E. Mooij,
H. Pothier, D. Esteve, C. Urbina, and M.H. Devoret, Phys. Rev.
Lett. {\bf 64}, 2691 (1990).
\bibitem{Anatram}
M.P. Amatram and S. Datta, Phys. Rev. B {\bf 51},
7632 (1995).
\bibitem{Gorelik97}
L.Y. Gorelik, F.A. Maa\o\, R.I. Shekhter, and M. Jonson, Phys.
Rev. Lett. {\bf 78}, 3169 (1997).
\bibitem{Gorelik01}
S. Blom and L.Y. Gorelik, Phys. Rev. B {\bf 64}, 045320 (2001).
\bibitem{Zozou99}
I.V. Zozoulenko, A.S. Sachrajda, C. Gould, K.-F. Berggrem, P.
Zawadzki, Y. Feng, and Z. Wasilewski, Phys. Rev. Lett. {\bf 83},
1838 (1999).


\bibitem{Smith} F.T. Smith, Phys. Rev. {\bf 118}, 349 (1960).


\bibitem{Qin01} H. Qin, F. Simmel, R.H. Blick, J.P. Kotthaus, W.
      Wegscheider, and M. Bichler, Phys. Rev. B {\bf 63}, 035320 (2001).


\bibitem{tang-unp} C.S. Tang and C.S. Chu (unpublished).

\end{references}
\end{document}